\documentclass[11pt,a4paper]{article}

\usepackage{amsmath,amssymb,amsthm}

{\theoremstyle{definition}

\newtheorem*{note*}{Note}
}
\begin{document}
\par\noindent {\LARGE\bf
Multidimensional simple waves \\
in fully relativistic fluids

\par}

{\vspace{5mm}\par\noindent {\it TALIYA SAHIHI~$^\dag$, HOMAYOON
ESHRAGHI~$^\dag$ and ALI MAHDIPOUR--SHIRAYEH~$^\ddag$}
\par\vspace{2mm}\par}

{\vspace{2mm}\par\noindent \it $^\dag$\,\, School of Physics, Iran
University of Science and Technology (IUST),
\\
$\phantom{\dag}$\,\,Narmak, Tehran, Iran}

{\vspace{2mm}\par\noindent \it $^\ddag$\,\, School of Mathematics,
Iran University of Science and Technology (IUST),
\\
$\phantom{\dag}$\,\,Narmak, Tehran, Iran}

{\vspace{2mm}\par\noindent \rm
E-mail:\it$^\dag$taliasah@yahoo.com,eshraghi@iust.ac.ir,
$^\ddag$mahdipour@iust.ac.ir}

{\vspace{8mm}\par\noindent\hspace*{10mm}\parbox{140mm}{\small A
special version of multi--dimensional simple waves given in
[G.~Boillat, {\it J. Math. Phys.} {\bf 11}, 1482-3 (1970)] and
[G.M.~Webb, R.~Ratkiewicz, M.~Brio and G.P.~Zank, {\it J. Plasma
Phys.} {\bf 59}, 417-460 (1998)] is employed for  fully relativistic
fluid and plasma flows. Three essential modes: vortex, entropy and
sound modes are derived where each of them is different from its
nonrelativistic analogue. Vortex and entropy modes are formally
solved in both the laboratory frame and the wave frame (co-moving
with the wave front) while the sound mode is formally solved only in
the wave frame at ultra-relativistic temperatures. In addition, the
surface which is the boundary between the permitted and forbidden
regions of the solution is introduced and determined. Finally a
symmetry analysis is performed for the vortex mode equation up to
both point and contact transformations. Fundamental invariants and a
form of general solutions of point transformations along with some
specific examples are also derived. }\par\vspace{6mm}}

\vspace{5mm} \hspace{2mm} Key words: Relativistic Fluids,
Multidimensional simple waves, Symmetry analysis


\section{Introduction}
Investigation of nonlinear phenomena appearing in a very wide area
of pure and applied sciences has met extremely extensive progresses
and developments. These studies which split into numerical and
analytical considerations are essentially and in most cases related
to some nonlinear differential equations. In spite of such a huge
amount of improvements almost all of these equations are still far
from being understood well. Among those nonlinear systems, a few
interesting open problems concern the hydrodynamic type of equations
governing fluid motions. Especially the Euler and Navier-Stokes
equations which reveal a mysterious behavior  are being intensively
studied in two main considerations: The incompressible motion mostly
dealing with vortex dynamics and the compressible flow concerning
the appearance of discontinuities shocks. Both of these problems are
somehow related to the debate of the regularity of solutions. In the
former consideration the aim is to understand the mechanism of
occurrence of vortex singularities while in the latter the shock
convergence is the most important question\cite{smoller,Dafermos}.
The core of these subjects is the following principal open question:
Starting from an initially smooth flow how can we predict the
appearance of any kind of discontinuity or singularity in later
times? In other words, what is ``{\it hidden}" in the smooth initial
conditions which causes the non-smoothness in future? Numerical
evidences as well as some analytical investigations in special cases
highly confirm the existence of such hidden facts. It seems that
these facts are partly related to topological properties and partly
to the measure theoretical aspects.

The wide variety of the application of fluid motions causes to deal
with all kinds of differential equations namely, elliptic, parabolic
and hyperbolic equations. While many systematic treatments have been
found and developed for parabolic and elliptic equations, hyperbolic
equations are still out of the frame of any well-defined method.
These equations posses some characteristic curves or surfaces which
naturally have the capability of forming discontinuities. This of
course makes the nature of the solution to be ``local" which means
that the continuous solution may exist only in some part of the
space and in some intervals of time. Generally there are many
topological, geometrical and analytical unknown features determining
the validity of any solution which are very difficult to discover
under the present human knowledge and thus new tools are needed.

We believe the best way to obtain some result is to study useful
special cases guiding us to more general statements. An excellent
and rich class of solutions for compressible ideal flows governed by
hyperbolic equations lies in the framework of simple (Riemann) waves
which have the capability of shock formation\cite{smoller,Dafermos}
and blow up occurrence\cite{John}. Simple waves constructed on the
basis of characteristics are clearly local which were discovered
first by Riemann in the 1-D form\cite{VonMises}-\cite{Landau} and
are still the best analytical tools to achieve shock
waves\cite{smoller,Dafermos,Courant,Akhiezer,Stenflo,Shukla}. By
imposing more restrictions and limitations on these solutions it was
possible to build multidimensional simple
waves\cite{Burnat1}-\cite{Burnat3}. Even a more generalization
yields double waves and multi-waves which give more advanced
solutions with more intensive blow
up\cite{Burnat3}-\cite{Tskhakaya1}.

It is obvious that taking into account relativistic effects highly
increases the coupling and so the nonlinearity of fluid and plasma
motions. Relativistic flows have been known for a long
time\cite{Lichnerowicz1}-\cite{Anile} and especially they are
important is astrophysical and cosmological phenomena. In addition,
under recent technical progresses in laser-plasma interaction,
plasma accelerators and fusion plasmas, the access to relativistic
effects in the laboratory is now very easy. Therefore a great
attention has been paid to analyze relativistic flows. Again the
study of simple waves plays a very fundamental role as almost the
only available nonstationary exact solution with the ability of
discontinuity formation.

A very excellent and complete mathematical discussion on one
dimensional relativistic MHD simple waves has been reviewed by
Shikin\cite{Shikin}. Some solutions of these 1-D simple waves are
found in many papers. Although a relativistic 2-D double wave
solution solution has been given only for ultra-relativistic
fluids\cite{Tskhakaya1} but still we observe the missing of a
multidimensional simple wave for a fully relativistic flow. This
task is the aim of the present note in which the approach of
Ref.~\cite{Webb2} is employed and generalized.

Physically it is a valid question that why we should consider
relativistic fluids while usually at so high temperatures matter is
found in the plasma form and so one has to take into account
electromagnetic fields leading to MHD equations. However this is not
always true because sometimes we deal with neutral fluids like
neutron stars. Moreover in the absence of any external magnetic
field and when the typical length and time for the non-neutrality of
the plasma are sufficiently less than the length and time for
macroscopic motions, the plasma can be considered as a neutral fluid
with at a very high accuracy. Hence it has sense to consider the
ideal relativistic flow here.

In the next section after a brief derivation of relativistic ideal
fluid equations, a multidimensional simple wave ansatz is
substituted into these equations and various modes and phase
velocities relative to the laboratory (fixed) frame are found. In
Sec.~3 some solutions for the vortex and entropy modes are given
only in the laboratory frame. The presented solutions are very
general and formal and a detailed solution is very difficult and
needs to determine the initial and boundary conditions precisely.
Thus, our solutions are very general including many arbitrary
functions. In Sec.~4 the equations are rewritten in the wave frame
and again some simple typical solutions are given for the all three
modes vortex, entropy and sound. Especially for the sound since its
equations are very complicated in the laboratory frame, it is seen
in Sec.~4 that these equations in the wave frame at
ultra-relativistic temperatures are simplified and it will be
possible to obtain some formal solutions for it. In sec.~5 we
investigate symmetry properties and their related topics for the
vortex mode equation as a sample equation appearing in our problem.
Finally a summary and concluding remarks are given in Sec.~6.

\section{Multidimensional simple wave formulation}

Relativistic effects in continuum matters in two aspects: Large
macroscopic (fluid) velocities and relativistic temperatures at
which the mean thermal energy of particles are comparable with their
rest energy. Both of these aspects are included in the
energy-momentum tensor
\begin{equation}\label{1}
T^k_i =wu^ku_i -P\delta^k_i~, \;\;\;\quad (i,k=0,1,2,3)
\end{equation}
where $u^j=(\gamma , \gamma\mathbf{v}/c)$ is the contra-variant
4-velocity and thus $u_j=(\gamma , -\gamma\mathbf{v}/c)$ is the
co-variant 4-velocity and $w=\varepsilon +P$ in which $P$ is the
fluid pressure and $\varepsilon $ is the internal energy (including
the rest energy) per unit proper volume (unit volume in the inertial
frame in which the fluid is momentarily at rest). Therefore $w$ is
the enthalpy per unit proper volume. Also $\mathbf{v}$ is the fluid
velocity and $\gamma =(1-v^2/c^2)^{-1/2}$ where $c$ is the speed of
light.

Basic equations consist of continuity equation
\begin{equation}\label{2}
\frac{\partial}{\partial x^i}(nu^i)=0~, \quad\quad \texttt{or}
\quad\quad \frac{1}{c}\frac{\partial}{\partial t}(\gamma n) +
\boldsymbol{\nabla}\cdot (n\gamma \mathbf{v})=0~,
\end{equation}
and the vanishing 4-divergence of the energy-momentum tensor
\begin{equation}\label{3}
\frac{\partial}{\partial x^k}T^k_i =0~, \quad\quad\quad (i=0,1,2,3)
\end{equation}
where $n$ is the number density of fluid particles in the proper
frame. By virtue of thermodynamic identity $TdS=d(w/n)-dP/n$ ($T$ is
the fluid temperature and $S$ is the entropy per unit particle) one
can combine Eqs.~(\ref{2}) and (\ref{3}) to obtain\cite{Shikin}
\begin{equation}\label{4}
\frac{dS}{dt}=\frac{\partial S}{\partial t} + \mathbf{v}\cdot
\boldsymbol{\nabla}S=0~.
\end{equation}
This equation can be alternatively considered in place of the zeroth
component ($i=0$) of Eq.~(\ref{3}). Thus our set of equations
consists of five equations (\ref{2}), (\ref{4}) and the spatial
components of (\ref{3}). This system of course needs a
thermodynamical state equation $P=P(S,w)$. However it is found that
this system of equations takes a more appropriate form by the use of
the following useful transformation\cite{Shikin}
\begin{equation}\label{5}
\kappa^i =\frac{w}{mnc}u^i \quad\quad ~,\quad\quad
\tilde{\rho}=\frac{(mnc)^2}{w}~,
\end{equation}
where $m$ is the ``mean" rest mass of all particles in the fluid.
This transformation makes our final system of equations to
\begin{equation}\label{6}
\frac{1}{c}\frac{\partial}{\partial t}(\tilde{\rho}\kappa_0) +
\boldsymbol{\nabla}\cdot (\tilde{\rho}\boldsymbol{\kappa})=0~,
\end{equation}
\begin{equation}\label{7}
\frac{\kappa_0}{c}\frac{\partial \boldsymbol{\kappa}}{\partial t} +
(\boldsymbol{\kappa}\cdot\boldsymbol{\nabla})\boldsymbol{\kappa} =-
\frac{1}{\tilde{\rho}}\boldsymbol{\nabla}P~,
\end{equation}
\begin{equation}\label{8}
\frac{\kappa_0}{c}\frac{\partial S}{\partial t} +
\boldsymbol{\kappa}\cdot\boldsymbol{\nabla}S =0~,
\end{equation}
\begin{equation}\label{9}
P=P(S,\tilde{\rho})~.
\end{equation}
Here $\kappa^i=(\kappa^0,\boldsymbol{\kappa})$ and thus
$\kappa_i=(\kappa_0,-\boldsymbol{\kappa})$ and
\begin{equation}\label{10}
\kappa^0=\kappa_0=\sqrt{\kappa^2 + w/\tilde{\rho}}~,\quad\quad
(\kappa =|\boldsymbol{\kappa}|)
\end{equation}
which follows from the identity $u^iu_i=1$.

The special feature of a simple wave in any quasi-linear hyperbolic
system of equations is that all quantities are considered as
functions of only one variable which we call it the phase and denote
by $\varphi =\varphi (\mathbf{r},t)$. In our problem we write
\begin{equation}\label{11}
\mathbf{U}=\mathbf{U}(\varphi)~,
\end{equation}
where
\begin{equation}\label{12}
\mathbf{U}=(\tilde{\rho},\kappa_1,\kappa_2,\kappa_3,S)
\end{equation}
is the state vector of the system. Boillat\cite{Boillat,Webb2}
showed that for a simple wave it is necessary to have
\begin{equation}\label{13}
\frac{\boldsymbol{\nabla}\varphi}{|\boldsymbol{\nabla}\varphi
|}\equiv \mathbf{n}=\mathbf{n}(\varphi)\quad\quad , \quad\quad
-\frac{\partial\varphi /\partial t}{|\boldsymbol{\nabla}\varphi
|}\equiv\lambda =\lambda (\varphi)~.
\end{equation}
In other words, the unit vector $\mathbf{n}$ normal to the wave
front must be only a function of $\varphi$ and the same is true for
the phase velocity $\lambda$. Finally condition (\ref{13}) implies
that $\varphi$ must satisfy\cite{Boillat,Webb2}
\begin{equation}\label{14}
G(\varphi,\mathbf{r},t)=f(\varphi) + \lambda (\varphi)t -
\mathbf{r}\cdot \mathbf{n}(\varphi)~,
\end{equation}
in which $f$ is an arbitrary differentiable function to be fixed
through initial conditions. This equation clearly means that level
surfaces of $\varphi$ are flat planes. The functional form of
$\mathbf{n}(\varphi)$ can not be determined from any equation and so
it remains arbitrary to be flexible to fit with a given condition.

There are two noticeable points about these multidimensional simple
waves. The first is the wave breaking at which the time and spatial
derivatives of $\varphi$ and so all variables diverge when
$F\longrightarrow 0$ provided that
\begin{equation}\label{15}
\frac{\partial \varphi}{\partial
t}=-\frac{\lambda(\varphi)}{F}\quad\quad , \quad\quad
\boldsymbol{\nabla}\varphi =\frac{\mathbf{n}(\varphi)}{F}~,
\end{equation}
\begin{equation}\label{16}
F\equiv \frac{\partial
G}{\partial\varphi}=\frac{df(\varphi)}{d\varphi} +
\frac{d\lambda(\varphi)}{d\varphi}t - \mathbf{r}\cdot
\frac{d\mathbf{n}(\varphi)}{d\varphi}= \frac{1}{|
\boldsymbol{\nabla}\varphi |}~.
\end{equation}
Equations (\ref{15}) are easily derived by implicit time and space
differentiations of (\ref{14}). Thus our simple wave solution is
valid only when $F>0$ and at any time and point where $F=0$ the
solution is not correct. The second point arises from the dependence
of $\mathbf{n}$ on $\varphi$ which implies that for two different
values $\varphi_1$ and $\varphi_2$ of $\varphi$ generally
$\mathbf{n}(\varphi_1)$ and $\mathbf{n}(\varphi_2)$ are not parallel
and thus they have an intersection on a line at which  the solution
is multi-valued which is not accepted. Hence, the domain of the
valid solution must not contain such intersections. Both of these
points demonstrate the ``local" character of simple waves.

For a unidirectional 1-D simple wave where $\mathbf{n}$ is a
constant vector it is possible for each value of $\varphi$ to
calculate the time of wave breaking ($F=0$) as
$t_c(\varphi)=-(df/d\varphi)/d\lambda /d\varphi$ and the earliest
time of the wave breaking is obtained by solving the equation
$(dt_c/d\varphi)=0$\cite{Akhiezer}. Unfortunately such a nice
situation does not hold in the multidimensional case when
$\mathbf{n}=\mathbf{n}(\varphi)$. Let us see this in a quantitative
way. Singular points (wave breaking) not only must satisfy the
simple wave condition (\ref{14}) but also they should fulfil
\begin{equation}\label{17}
F=0~.
\end{equation}
Thus, the wave breaking occurs on the line of intersection of the
two perpendicular planes $G=0$ and $F=0$. This line is exactly the
rotation axis of the wave front at $\varphi$ when $\varphi$ has an
infinitesimal growth to $\varphi + \delta\varphi$. This will be
easily seen if we observe that the wave front for $\varphi +
\delta\varphi$ must satisfy
$$
G(\varphi + \delta\varphi,\mathbf{r},t)=0~, \quad\quad
\texttt{or}\quad\quad G(\varphi ,\mathbf{r},t)+F\delta\varphi=0~,
$$
which again yields Eqs.~(\ref{14}) and (\ref{17}). We may therefore
conclude (without a rigorous proof) that if the wave breaking
(singularity) line lies out of the region of the solution, the line
of multi-valuedness will also lay in that region. Besides, since a
line of singularity for each value of $\varphi$ exists at each
instant of time, it has no sense to speak about $t_c(\varphi)$.
However if the fluid fills the whole space $\mathbb{R}^3$ we can
obtain a moving surface constructed at any time exactly from all of
these singular lines at that time. This surface is in fact the
boundary between the forbidden and permitted regions relative to a
simple wave solution.

Now we substitute the simple wave ansatz (\ref{11}) into
Eqs.~(\ref{6})-(\ref{8}) supplemented by Eq.~(\ref{9}) and then
divide each equation by $\mid \boldsymbol{\nabla\varphi} \mid$ and
use (\ref{13}) to obtain the system of five quasi-linear coupled
equations
\begin{equation}\label{18}
A~\frac{d\mathbf{U}}{d\varphi}=0~,
\end{equation}
where $A$ is the $5\times 5$ matrix with the following elements
\begin{equation}\label{19}
A= \left [ \begin{array}{ccccc}
\kappa_n-\frac{\lambda}{c}(\tilde{\rho}\frac{\partial\kappa_0}
{\partial\tilde{\rho}}+\kappa_0) &
\tilde{\rho}(n_1-\frac{\lambda}{c}\frac{\kappa_1}{\kappa_0})
&\tilde{\rho}(n_2-\frac{\lambda}{c}\frac{\kappa_2}{\kappa_0})
&\tilde{\rho}(n_3-\frac{\lambda}{c}\frac{\kappa_3}{\kappa_0})
&-\frac{\lambda}{c}\tilde{\rho}\frac{\partial\kappa_0}
{\partial S}  \\
\frac{a^2n_1}{\tilde{\rho}} & \kappa_n- \frac{\lambda}{c}\kappa_0
& 0 & 0 & \frac{P_Sn_1}{\tilde{\rho}} \\
\frac{a^2n_2}{\tilde{\rho}} & 0 &\kappa_n- \frac{\lambda}{c}\kappa_0
& 0 & \frac{P_Sn_2}{\tilde{\rho}} \\
\frac{a^2n_3}{\tilde{\rho}} & 0 &0 & \kappa_n- \frac{\lambda}{c}\kappa_0
& \frac{P_Sn_2}{\tilde{\rho}} \\
0 & 0 & 0 & 0 & \kappa_n- \frac{\lambda}{c}\kappa_0 \\
\end{array}
\right].
\end{equation}
In the above matrix we have used the following notations
\begin{equation}\label{20}
\kappa_n \equiv \boldsymbol{\kappa}\cdot
\mathbf{n}=\sum_{i=1}^{3}\kappa_in_i\quad ,\quad a^2\equiv \left(
\frac{\partial P}{\partial \tilde{\rho}}\right)_S\quad ,\quad
P_S\equiv \left(\frac{\partial P}{\partial
S}\right)_{\tilde{\rho}}~.
\end{equation}
Moreover, in the calculation of $\frac{\partial \kappa_0}{\partial
\tilde{\rho}}$ and $\frac{\partial \kappa_0}{\partial S}$ we must
assume $w=w(\tilde{\rho},S)$ and use Eq.~(\ref{10}) to express
$\kappa_0$ explicitly as a function of all five variables
$\mathbf{U}=(\tilde{\rho},\boldsymbol{\kappa},S)$.

Equation (\ref{18}) has a nontrivial solution only when
\begin{equation}\label{21}
\det(A)=0~,
\end{equation}
which constructs a fifth order equation for $\lambda$ with a triple
root
\begin{equation}\label{22}
\lambda_1=\lambda_2=\lambda_3=c\frac{\kappa_n}{\kappa_0}=v_n~,
\end{equation}
while the fourth and fifth roots $\lambda_4$ and $\lambda_5$ are the
larger and smaller roots of the following quadratic equation
respectively.
\begin{equation}\label{23}
\left[\kappa_n-\frac{\lambda}{c}\left(\tilde{\rho}\frac{\partial\kappa_0}
{\partial\tilde{\rho}}+\kappa_0\right)\right]\left(\kappa_n-\frac{\lambda}{c}
\kappa_0\right)=a^2\left(1-\frac{\lambda}{c}\frac{\kappa_n}{\kappa_0}\right)~.
\end{equation}
The triplet root is the phase velocity for the two vortex modes and
one entropy to be discussed in the next section. The roots
$\lambda_4$ and $\lambda_5$ are the phase velocities for the forward
and backward sound modes respectively. Although these modes have
nonrelativistic analogue but they significantly differ from the
nonrelativistic case.

Substitution of each value of the phase velocity into (\ref{18})
yields some ordinary differential equations for
$\mathbf{U}(\varphi)$ to be solved. For the entropy and vortex modes
these equations are not difficult and some formal solutions both in
the laboratory frame and the wave frame will be presented in
Sections 3 and 4 respectively. Since the equations for the sound
waves are complicated in the laboratory frame, we go to the wave
frame but still they are difficult to solve and finally when we
consider the physically common case of ultra-relativistic
temperatures it will be possible to obtain some solutions in Sec.~4.

\section{Vortex and entropy modes}

If we substitute the triplet root
$\lambda=c\frac{\kappa_n}{\kappa_0}$ into (\ref{18}) we obtain
\begin{equation}\label{24}
-\frac{\kappa_n}{\kappa_0}\frac{\partial\kappa_0}{\partial\tilde{\rho}}
\frac{d\tilde{\rho}}{d\varphi}+\left(
\mathbf{n}-\frac{\kappa_n}{\kappa_0^2}\boldsymbol{\kappa}\right)
\cdot\frac{d\boldsymbol{\kappa}}{d\varphi}~-\frac{\kappa_n}{\kappa_0}
\frac{\partial \kappa_0}{\partial S}\frac{dS}{d\varphi}=0~,
\end{equation}
\begin{equation}\label{25}
a^2\frac{d\tilde{\rho}}{d\varphi} +
P_S\frac{dS}{d\varphi}=\frac{dP}{d\varphi}=0\Longrightarrow
P(\varphi)=\texttt{const}~,
\end{equation}
\begin{equation}\label{26}
0\cdot \frac{dS}{d\varphi}=0~.
\end{equation}
In Eq.~(\ref{25}) we have used the second and third equations of
(\ref{20}) together with (\ref{9}). Equation (\ref{26}) admits the
two cases of constant entropy (vortex modes) and variable entropy
(entropy mode).

\subsection{Vortex modes}

We have $dS=0$ or
\begin{equation}\label{27}
S(\varphi)=\mbox{const}~,
\end{equation}
which together with (\ref{25}) and the state equation (\ref{9})
yields the constancy of $\tilde{\rho}$ and so all thermodynamical
variables. Thus, only the fluid velocity considered in
$\boldsymbol{\kappa}$ and $\mathbf{n}$ change with $\varphi$ which
$\mathbf{n}(\varphi)$ is an arbitrary suitable function. Regarding
the above results in Eqs.~(\ref{24}) and (\ref{10}) one can obtain
the equation for $\boldsymbol{\kappa}(\varphi)$.
\begin{equation}\label{28}
\left(
\mathbf{n}-\frac{\kappa_n}{\kappa_0^2}\boldsymbol{\kappa}\right)\cdot
\frac{d\boldsymbol{\kappa}}{d\varphi}=0~,
\end{equation}
which must be supplemented by
\begin{equation}\label{29}
\kappa_0=\sqrt{\kappa^2 + w_0/\tilde{\rho}}_0~,
\end{equation}
where $w_0$ and $\tilde{\rho}_0$ are constant throughout the wave.
The factor $\left(\mathbf{n}-\frac{\kappa_n}{\kappa_0^2}
\boldsymbol{\kappa}\right)$ in (\ref{28}) can not be zero because if
it is zero we can make its inner product  with $\mathbf{n}$ and
obtain $\kappa_n^2=\kappa^2=\kappa_0^2$ which is impossible by
(\ref{29}). Therefore Eq.~(\ref{28}) is equivalent to
\begin{equation}\label{30}
\frac{d\boldsymbol{\kappa}}{d\varphi}=\mathbf{X}(\varphi)\times
\left(\mathbf{n}-\frac{\kappa_n}{\kappa_0^2}
\boldsymbol{\kappa}\right)~,
\end{equation}
where $\mathbf{X}(\varphi)$ is an arbitrary continuous function. It
is possible to choose two functions  $\mathbf{X}_1(\varphi)$ and
$\mathbf{X}_2(\varphi)$ where $\mathbf{X}_1(\varphi)\cdot
\mathbf{X}_2(\varphi)=0$ which gives two perpendicular and
independent vortex modes similar to the nonrelativistic
case\cite{Webb2}.

It is also worth noting that we can define a generalized vortex
\begin{equation}\label{31}
\boldsymbol{\Omega}\equiv \boldsymbol{\nabla}\times
\boldsymbol{\kappa} = \boldsymbol{\nabla}\varphi \times
\frac{d\boldsymbol{\kappa}}{d\varphi}=|
\boldsymbol{\nabla}\varphi|\mathbf{n}\times
\frac{d\boldsymbol{\kappa}}{d\varphi}~,
\end{equation}
which is constant not only on the wave front but also in advection
with the fluid velocity:
\begin{equation}\label{32}
\frac{\partial\boldsymbol{\Omega}}{\partial t} + (\mathbf{v}\cdot
\boldsymbol{\nabla})\boldsymbol{\Omega} =|
\boldsymbol{\nabla}\varphi |\left( \frac{c\kappa_n}{\kappa_0}
-\lambda \right)\frac{d\boldsymbol{\Omega}}{d\varphi}=0~,
\end{equation}
This fact is consistent with the ``frozen in" condition of filed
lines of $\boldsymbol{\Omega}$\cite{Eshraghi}
\begin{equation}\label{33}
\frac{\partial}{\partial t}\left(\frac{\boldsymbol{\Omega}}{\gamma
n}\right)+(\mathbf{v}\cdot
\boldsymbol{\nabla})\left(\frac{\boldsymbol{\Omega}}{\gamma
n}\right)=\left(\frac{\boldsymbol{\Omega}}{\gamma
n}\cdot\boldsymbol{\nabla}\right)\mathbf{v}~,
\end{equation}
in which the number density in the laboratory frame $\gamma n$ is
constant because
\begin{equation}\label{34}
\boldsymbol{\nabla}\cdot \mathbf{v}=c\boldsymbol{\nabla}\cdot \left(
\frac{\boldsymbol{\kappa}}{\kappa_0}\right)=c\frac{|\boldsymbol{\nabla}\varphi|}
{\kappa_0}\left(
\mathbf{n}-\frac{\kappa_n}{\kappa_0^2}\boldsymbol{\kappa}\right)\cdot
\frac{d\boldsymbol{\kappa}}{d\varphi}=0~,
\end{equation}
according to Eq.~(\ref{28}). Finally by (\ref{31}) we find
$$
\boldsymbol{\Omega}\cdot\boldsymbol{\nabla}=
|\boldsymbol{\nabla}\varphi|\boldsymbol{\Omega}\cdot
\mathbf{n}\frac{d}{d\varphi}=0~,
$$
by which Eq.~(\ref{33}) reduces to (\ref{32}).

Equation (\ref{30}) has many solutions since $\mathbf{X}(\varphi)$
is an arbitrary continuous function. It is therefore easy to choose
some suitable simple forms for $\mathbf{X}$ such that Eq.~(\ref{30})
can be easily solved. As an example consider two different forms
\begin{equation}\label{35}
\mathbf{X}=\alpha \boldsymbol{\kappa}~, \quad\quad \mbox{or
alternatively} \quad\quad \mathbf{X}=\alpha
\frac{\kappa_0^2}{\kappa_n}\mathbf{n}~,
\end{equation}
where $\alpha$ is a dimensionless constant. Selecting each value
form for $\mathbf{X}$ from (\ref{35})causes to reduce Eq.~(\ref{30})
to
\begin{equation}\label{36}
\frac{d\boldsymbol{\kappa}}{d\varphi}=\alpha \boldsymbol{\kappa}
\times \mathbf{n}(\varphi)~.
\end{equation}
Regardless of the functional form of $\mathbf{n}(\varphi)$ it is
obvious from (\ref{36}) that $|\boldsymbol{\kappa}(\varphi)|$ is
constant and from (\ref{29}) $\kappa_0$ is also constant and only
the direction of $\boldsymbol{\kappa}(\varphi)$ changes by
$\varphi$. To obtain a more special solution let us consider a 2D
simple wave with\cite{Webb2}
\begin{equation}\label{37}
\mathbf{n}(\varphi)=(-\sin\varphi , \cos\varphi , 0)~,
\end{equation}
in a Cartesian coordinate system. In this case we have
\begin{equation}\label{38}
\boldsymbol{\kappa}=\kappa_n\mathbf{n} + \kappa_t\mathbf{t} +
\kappa_3\mathbf{z}~,
\end{equation}
where
\begin{equation}\label{39}
\kappa_n=\boldsymbol{\kappa}\cdot\mathbf{n}=-\kappa_1\sin\varphi +
\kappa_2\cos\varphi \quad\quad , \quad\quad
\kappa_t=\boldsymbol{\kappa}\cdot\mathbf{t}= -(\kappa_1\cos\varphi +
\kappa_2\sin\varphi)~,
\end{equation}
in which $\mathbf{t}=(-\cos\varphi , -\sin\varphi , 0)$ is normal to
$\mathbf{n}$. Substitution of (\ref{38}) into (\ref{36}) and using
the relations
\begin{equation}\label{40}
\frac{d\mathbf{n}}{d\varphi}=\mathbf{t}\quad ,\quad
\frac{d\mathbf{t}}{d\varphi}=-\mathbf{n}\quad ,\quad
\mathbf{n}\times \mathbf{t}=\mathbf{z}~,
\end{equation}
one finds
\begin{gather}
\label{41}
\frac{d\kappa_n}{d\varphi}=\kappa_t~,
\\\label{42}
\frac{d\kappa_t}{d\varphi} + \kappa_n - \alpha\kappa_3=0~,
\\\label{43}
\frac{d\kappa_3}{d\varphi}=-\alpha \kappa_t~.
\end{gather}

Equations (\ref{41}) and (\ref{43}) yield
\begin{equation}\label{44}
\kappa_3=-\alpha\kappa_n + \bar{\kappa}~,
\end{equation}
where $\bar{\kappa}$ is a constant with the dimension of $\kappa$
(velocity). Then we substitute $\kappa_t$ from (\ref{41}) and
$\kappa_3$ from (\ref{44}) into (\ref{42}) to obtain
\begin{equation}\label{45}
\frac{d^2\kappa_n}{d\varphi^2} + (1+\alpha^2)\kappa_n - \alpha
\bar{\kappa}=0~,
\end{equation}
with a general solution
\begin{equation}\label{46}
\kappa_n =\bar{\kappa}_n\cos[\sqrt{1+\alpha^2}(\varphi + \beta)]
+\frac{\alpha}{1+\alpha^2}\bar{\kappa}~,
\end{equation}
where $\bar{\kappa}_n$ is a constant with the dimension of $\kappa$
(velocity) while $\beta$ is a dimensionless constant. Then from
(\ref{41}) we have
\begin{equation}\label{47}
\kappa_t=-\sqrt{1+\alpha^2}\bar{\kappa}_n\sin[\sqrt{1+\alpha^2}(\varphi
+ \beta)]~,
\end{equation}
and from (\ref{44}) we find
\begin{equation}\label{48}
\kappa_3 =-\alpha\bar{\kappa}_n\cos[\sqrt{1+\alpha^2}(\varphi +
\beta)] +\frac{1}{1+\alpha^2}\bar{\kappa}~,
\end{equation}
It is then straightforward to find $\kappa_1$ and $\kappa_2$ by the
use of
\begin{equation}\label{49}
\kappa_1=-(\kappa_n\sin\varphi + \kappa_t\cos\varphi)\quad \quad ,
\quad \quad \kappa_2=\kappa_n\cos\varphi - \kappa_t\sin\varphi~.
\end{equation}
As mentioned before a complete solution needs more detailed
informations about initial and boundary conditions which is not of
our interest here.

\subsection{Entropy modes}

Here we have $dS\neq 0$ in (\ref{26}) and thus $\tilde{\rho}$ is not
constant although by (\ref{25}) $P$ is still constant. It is
possible to reduce Eq.~(\ref{24}) to
\begin{equation}\label{50}
\mathbf{n}\cdot
\left(\frac{d\boldsymbol{\kappa}}{d\varphi}-\frac{d(\ln\kappa_0)}{d\varphi}
\boldsymbol{\kappa}\right)=0~,
\end{equation}
which gives
\begin{equation}\label{51}
\frac{d\boldsymbol{\kappa}}{d\varphi}=\mathbf{Y}(\varphi)\times\mathbf{n}
+\frac{d(\ln\kappa_0)}{d\varphi} \boldsymbol{\kappa}~,
\end{equation}
where $\mathbf{Y}(\varphi)$ is again an arbitrary continuous
function. Let us again choose a suitable form for
$\mathbf{Y}(\varphi)$ to simplify the solution. For example if
$\mathbf{Y}$ is parallel to $\mathbf{n}$ we see from (\ref{51}) that
\begin{equation}\label{52}
\boldsymbol{\kappa}=\kappa_0\mathbf{C}~,
\end{equation}
where $\mathbf{C}=(c_1,c_2,c_3)$ is a  dimensionless constant
vector. Since $P$ is constant, the enthalpy $w$ becomes only a
function of $\tilde{\rho}$ and thus Eqs.~(\ref{52}) and (\ref{10})
yield
\begin{equation}\label{53}
\boldsymbol{\kappa} =\frac{\mathbf{C}}{\sqrt{1-|\mathbf{C}|^2}}
\sqrt{\frac{w(\tilde{\rho})}{\tilde{\rho}}}~,
\end{equation}
which is valid if $|\mathbf{C}|<1$. It remains to specify the
dependence of $\tilde{\rho}$ on $\varphi$. This dependence is
arbitrary because the fixing of $\varphi$ is under our control and
we can assume it as an arbitrary function of one or more physical
variables\cite{Webb2}.

\section{Simple waves presented in the wave frame}

Sometimes apparent forms of those equations concerning a simple wave
solution are so complicated and not soluble easily. A mathematical
trick here is to rewrite all equations in terms of physical
variables as measured in the wave frame. A wave frame depends on the
special value of the phase $\varphi$ that is, for any value of
$\varphi$ there is a plane wave front defined by Eq.~(\ref{14})
moving with the phase velocity
$\mathbf{V}_{ph}=\lambda(\varphi)\mathbf{n}(\varphi)$ and we
consider a Lorentz transformation from the laboratory frame to the
frame co-moving with this wave front. It is thus clear that to each
value of $\varphi$ corresponds a unique wave frame.

We denote all quantities in the wave frame by a prime except scalar
quantities such as $\tilde{\rho},w,n$ etc which are either Lorentz
invariant or defined in the proper frame co-moving with the fluid.
Therefore for the 4-vector $\kappa^i$ we have
\begin{equation}\label{54}
\begin{array}{l}
\kappa_0=\cosh\xi\kappa'_0 + \sinh\xi\kappa'_n~, \\
\\
\kappa_n=\cosh\xi\kappa'_n + \sinh\xi\kappa'_0~, \\
\\
\boldsymbol{\kappa}_{\perp}=\boldsymbol{\kappa}'_{\perp}~,
\end{array}
\end{equation}
in which $\kappa'_n=\boldsymbol{\kappa}'\cdot \mathbf{n}$~,
$\boldsymbol{\kappa}'_{\perp}=\boldsymbol{\kappa}'-\kappa'_n
\mathbf{n}$ and $\xi$ depends on $\varphi$ thorough
\begin{equation}\label{55}
\tanh \xi=\frac{\lambda (\varphi)}{c}~.
\end{equation}
In the following subsections we apply this method for the vortex,
entropy and sound modes and give simple formal solutions for each
one.

\subsection{Vortex modes in the wave frame}

For this mode we already had
$\frac{\lambda}{c}=\frac{\kappa_n}{\kappa_0}$ which by the
substitution from (\ref{54}) and (\ref{55}) into it we find
\begin{equation}\label{56}
\kappa'_n =0~.
\end{equation}
It is therefore seen that in the wave frame the phase velocity takes
the simple form through the above equation. Then we substitute
(\ref{54}) and (\ref{55}) into (\ref{28}) and use (\ref{56}) and the
identity $\boldsymbol{\kappa}'_{\perp}\cdot \mathbf{n}=0$ to obtain
\begin{equation}\label{57}
\cosh^2\xi \mathbf{n}\cdot
\frac{d\boldsymbol{\kappa}'_{\perp}}{d\varphi}
-\frac{\sinh\xi}{\kappa'_0}\boldsymbol{\kappa}'_{\perp}\cdot
\frac{d\boldsymbol{\kappa}'_{\perp}}{d\varphi}+\sinh\xi\frac{d\kappa'_0}{d\varphi}
+\kappa'_0\cosh\xi\frac{d\xi}{d\varphi}=0~.
\end{equation}
Since $(\kappa_0,\boldsymbol{\kappa})$ is a 4-vector, Eq.~(\ref{29})
is invariant and due to (\ref{56})we have
\begin{equation}\label{58}
\kappa'_0=\sqrt{\kappa'^2_{\perp}+w_0/\tilde{\rho}_0}~,
\end{equation}
by which Eq.~(\ref{57}) reduces to
\begin{equation}\label{59}
\boldsymbol{\kappa}'_{\perp}\cdot\frac{d\mathbf{n}}{d\varphi}=
-\mathbf{n}\cdot\frac{d\boldsymbol{\kappa}'_{\perp}}{d\varphi}=
\frac{\kappa'_0}{\cosh\xi}\frac{d\xi}{d\varphi}~.
\end{equation}
Assuming $\mathbf{n}(\varphi)$ is a known function,
$\boldsymbol{\kappa}'_{\perp}$ and $\xi$ should satisfy
Eq.~(\ref{59}) which obviously admits a large amount of freedom. As
a very simple solution let us assume the restriction
\begin{equation}\label{60}
\frac{d\boldsymbol{\kappa}'_{\perp}}{d\varphi}=
-\kappa'_{\perp}\mathbf{n}~,
\end{equation}
which says that $\kappa'_{\perp}$ and $\kappa'_0$ are both constant
and thus Eq.~(\ref{59}) can be easily solved to give
\begin{equation}\label{61}
\int^{\xi}_{\xi_0}\frac{d\xi'}{\cosh\xi'}=
\arctan(\sinh\xi)-\arctan(\sinh\xi_0)=
\frac{\kappa'_{\perp}}{\kappa'_0}(\varphi -\varphi_0)~.
\end{equation}
From the above solution one can find $\xi$ in terms of $\varphi$ by
which through Eq.~(\ref{55}) we have $\lambda(\varphi)$. For
$\boldsymbol{\kappa}'_{\perp}$ we formally solve Eq.~(\ref{60}):
\begin{equation}\label{62}
\boldsymbol{\kappa}'_{\perp}=-\kappa'_{\perp}
\int^{\varphi}_{\varphi_0}\mathbf{n}(\varphi')d\varphi' +
\boldsymbol{\kappa}'_{\perp}(\varphi_0)~.
\end{equation}
Here we gave a very restricted solution just as an example to show
the procedure of the solution. Depending on initial and boundary
conditions it is in principle possible to find more realistic
solutions although it seems to be very difficult.

\subsection{Entropy mode in the wave frame}

Since the phase velocity
$\frac{\lambda}{c}=\frac{\kappa_n}{\kappa_0}$ is the same as for the
vortex mode, Eq.~(\ref{56}) is again valid here and transforming
Eq.~(\ref{50}) in a manner similar to that performed for the vortex
mode we again obtain Eq.~(\ref{59}) for the entropy wave too but
here since $P$ is constant, $w$ is only a function of $\tilde{\rho}$
and thus
\begin{equation}\label{63}
\kappa'_0=\sqrt{\kappa'^2_{\perp}+w(\tilde{\rho})/\tilde{\rho}}~.
\end{equation}
Hence, the meaning of (\ref{59}) for the entropy mode is different
from this equation for the vortex mode. Again as a very restricted
simple solution we suggest Eqs.~(\ref{60}) and (\ref{62}) for
$\boldsymbol{\kappa}'_{\perp}(\varphi)$ resulting in the constancy
of $\kappa'_{\perp}$ and assume a given form for
$\tilde{\rho}(\varphi)$ by which from (\ref{63}) we have
$\kappa'_0(\varphi)$ as a known function of $\varphi$ and thus
Eq.~(\ref{59}) has the formal solution
\begin{equation}\label{64}
\arctan(\sinh\xi)-\arctan(\sinh\xi_0)= \kappa'_{\perp}
\int^{\varphi}_{\varphi_0}\frac{d\varphi'}{\kappa'_0(\varphi')}~.
\end{equation}

\subsection{Sound mode in the wave frame}

At first we see that in the laboratory reference frame there are
five equations included in (\ref{18}) but due to Eq.~(\ref{21}) we
have only four independent equations. The last (fifth) equation of
(\ref{18}) by the use of (\ref{19}) implies Eq.~(\ref{27}) also
valid for the sound mode which its substitution into the first four
equations of (\ref{18}) yields
\begin{equation}\label{65}
\left(\kappa_n-\frac{\lambda}{c}\kappa_0\right)\frac{d\tilde{\rho}}{d\varphi}
-\frac{\lambda}{c}\tilde{\rho}\frac{d\kappa_0}{d\varphi}+\tilde{\rho}
\mathbf{n}\cdot\frac{d\boldsymbol{\kappa}}{d\varphi}=0~,
\end{equation}
as the continuity equation and
\begin{equation}\label{66}
\left(\kappa_n-\frac{\lambda}{c}\kappa_0\right)\frac{d\boldsymbol{\kappa}}
{d\varphi}+\frac{a^2}{\tilde{\rho}}\frac{d\tilde{\rho}}{d\varphi}\mathbf{n}=0~,
\end{equation}
as the momentum equation in which $a^2$ is defined from (\ref{20}).
Equations (\ref{65}) and (\ref{66}) are four equations but only
three of them are independent while the phase velocity is $\lambda
=\lambda_4$ or $\lambda =\lambda_5$ which are the roots of the
quadratic equation (\ref{23}).

Let us rewrite Eqs.~(\ref{65}) and (\ref{66}) in terms of the wave
frame quantities through Eqs.~(\ref{54}) to find
\begin{equation}\label{67}
\kappa'_n \frac{d\tilde{\rho}}{d\varphi}+\tilde{\rho}\kappa'_0
\frac{d\xi}{d\varphi}+\tilde{\rho}\frac{d\kappa'_n}{d\varphi}+
\cosh\xi~\tilde{\rho}\mathbf{n}\cdot
\frac{d\boldsymbol{\kappa'}_{\perp}}{d\varphi}=0~,
\end{equation}
and
\begin{equation}\label{68}
\cosh\xi~\frac{a^2}{\tilde{\rho}\kappa'_n}\frac{d\tilde{\rho}}{d\varphi}
\mathbf{n}+\frac{d\boldsymbol{\kappa'}_{\perp}}{d\varphi}+\mathbf{n}
\frac{d}{d\varphi}(\kappa'_n\cosh\xi+\kappa'_0\sinh\xi)+
(\kappa'_n\cosh\xi+\kappa'_0\sinh\xi)\frac{d\mathbf{n}}{d\varphi}=0~,
\end{equation}
with only three independent equations.

It is also necessary to rewrite the quadratic equation (\ref{23})
(whose roots are the sound waves $\lambda_4$ and $\lambda_5$) in
terms of the wave frame quantities. Substitution of (\ref{54}) and
(\ref{55}) into (\ref{23}) yields
\begin{equation}\label{69}
\left(\kappa'_n-\tilde{\rho}\frac{\partial\kappa_0}{\partial\tilde{\rho}}
\sinh\xi\right)\kappa'_n
=a^2\frac{\kappa'_0}{\kappa'_0+\kappa'_n\tanh\xi}~.
\end{equation}
Now we notice that according to Eq.~(\ref{10}) we have
\begin{equation}\label{70}
\frac{\partial\kappa_0}{\partial\tilde{\rho}}
=\frac{1}{2\kappa_0}\frac{\partial}{\partial\tilde{\rho}}\left(
\frac{w}{\tilde{\rho}}\right)_S=\frac{1}{2}\frac{1}{\kappa'_0\cosh\xi+\kappa'_n\sinh\xi}
\frac{d}{d\tilde{\rho}}\left( \frac{w}{\tilde{\rho}}\right)~,
\end{equation}
where we have used Eq.~(\ref{27}) by which the entropy is constant
and thus $w$ is only a function of $\tilde{\rho}$. We then
substitute Eq.~(\ref{70}) into (\ref{69}) to find
\begin{equation}\label{71}
(\kappa_n^{'2}-a^2)\kappa'_0 +
\left[\kappa_n^{'2}-\frac{\tilde{\rho}}{2}\frac{\partial}{\partial\tilde{\rho}}
\left(\frac{w}{\tilde{\rho}}\right)_S\right]\kappa'_n\tanh\xi=0~,
\end{equation}
in which
\begin{equation}\label{72}
\kappa'_0=\sqrt{\kappa'^2+w/\tilde{\rho}}~.
\end{equation}
Equation (\ref{71}) determines the phase velocity $\tanh\xi$ in
terms of the physical variables measured in the wave frame but since
it is generally complicated depending on the explicit form of
$w(\tilde{\rho})$ we can not go further. However, it is possible to
continue for the ultra-relativistic case when $K_BT\gg mc^2$ which
implies that
\[
\frac{w}{w_{\circ}}=\left(\frac{T}{T_{\circ}}\right)^4\quad\quad ,
\quad\quad
\frac{n}{n_{\circ}}=\left(\frac{T}{T_{\circ}}\right)^3\quad\quad ,
\quad\quad
P=\frac{1}{4}w~,
\]
by which it is easy to see
\begin{equation}\label{73}
\frac{w}{\tilde{\rho}}=\frac{w_{\circ}}{\tilde{\rho}^2_{\circ}}
\tilde{\rho}\quad\quad\quad , \quad\quad\quad
a^2=\frac{dP}{d\tilde{\rho}}=\frac{1}{2}\frac{w_{\circ}}{\tilde{\rho}^2_{\circ}}
\tilde{\rho}~,
\end{equation}
where the subscript "$\circ$'' denotes the equilibrium point of the
fluid at which it is at rest. By the above simplifications
Eq.~(\ref{71}) reduces to
\[
(\kappa_n^{'2}-\frac{w_{\circ}}{2\tilde{\rho}^2_{\circ}}\tilde{\rho})
(\kappa'_0 + \kappa'_n\tanh\xi )=0~.
\]
Since $|\kappa'_0/\kappa'_n|>1$ while $|\tanh\xi |<1$, the second
factor in the above equation can not be zero and thus for the sound
mode in the ultra-relativistic case we obtain
\begin{equation}\label{74}
\kappa'_n=\mp a=\mp
\sqrt{\frac{w_{\circ}}{2\tilde{\rho}^2_{\circ}}\tilde{\rho}}
\quad\quad , \quad\quad \kappa'_0=\sqrt{\kappa_{\perp}^{'2}+
\frac{3}{2}\frac{w_{\circ}}{\tilde{\rho}^2_{\circ}}\tilde{\rho}}~.
\end{equation}
Here the upper (minus) sign indicates the case where the fluid
velocity is negative with respect to the wave front which means that
the wave runs faster than the fluid and thus it refers to the
forward sound wave. Similarly the lower (plus) sign refers to the
backward sound wave.

As mentioned before there are only three independent equations
namely Eq.~(\ref{68}) when Eq.~(\ref{74}) is substituted into it. It
is more convenient to to write Eq.~(\ref{68}) in the three
orthogonal directions $\mathbf{n}$, $\boldsymbol{\kappa'}_{\perp}$
and $d\mathbf{n}/d\varphi$. Thus, making the scalar product of
(\ref{68})[after the substitution of (\ref{74}) into it] by
$\mathbf{n}$ yields
\begin{gather}\label{75}
\left(\sinh\xi\sqrt{\frac{w_{\circ}}{2\tilde{\rho}^2_{\circ}}
\tilde{\rho}}~\mp\cosh\xi\sqrt{\kappa_{\perp}^{'2}+
\frac{3}{2}\frac{w_{\circ}}{\tilde{\rho}^2_{\circ}}\tilde{\rho}}~\right)
\left(\frac{3}{2}\sqrt{\frac{w_{\circ}}
{2\tilde{\rho}^2_{\circ}}}\frac{1}{\sqrt{\tilde{\rho}(\kappa_{\perp}^{'2}+
\frac{3}{2}\frac{w_{\circ}}{\tilde{\rho}^2_{\circ}}\tilde{\rho})}}
\frac{d\tilde{\rho}}{d\varphi}~\mp\frac{d\xi}{d\varphi}\right)\nonumber\\[1ex]\qquad
=-\frac{\sinh\xi}{\sqrt{\kappa_{\perp}^{'2}+\frac{3}{2}\frac{w_{\circ}}
{\tilde{\rho}^2_{\circ}}\tilde{\rho}}}\boldsymbol{\kappa'}_{\perp}\cdot
\frac{d\boldsymbol{\kappa'}_{\perp}}{d\varphi}+
\boldsymbol{\kappa'}_{\perp}\cdot \frac{d\mathbf{n}}{d\varphi}~,
\end{gather}
in which we have used the identity
$\boldsymbol{\kappa'}_{\perp}\cdot\mathbf{n}=0$ which gives
\begin{equation}\label{76}
\frac{d\boldsymbol{\kappa'}_{\perp}}{d\varphi}\cdot\mathbf{n}+
\boldsymbol{\kappa'}_{\perp}\cdot\frac{d\mathbf{n}}{d\varphi}=0~.
\end{equation}
Next, let us make the scalar product of (\ref{68}) by
$\boldsymbol{\kappa'}_{\perp}$:
\begin{equation}\label{77}
\boldsymbol{\kappa'}_{\perp}\cdot\frac{d\boldsymbol{\kappa'}_{\perp}}
{d\varphi}=-\left(\sinh\xi\sqrt{\kappa_{\perp}^{'2}+\frac{3}{2}\frac{w_{\circ}}
{\tilde{\rho}^2_{\circ}}\tilde{\rho}}~\mp\cosh\xi\sqrt{\frac{w_{\circ}}
{2\tilde{\rho}^2_{\circ}}\tilde{\rho}}~\right)\boldsymbol{\kappa'}_{\perp}
\cdot\frac{d\mathbf{n}}{d\varphi}~.
\end{equation}
Finally the scalar product of (\ref{68}) by $d\mathbf{n}/d\varphi$
is
\begin{equation}\label{78}
\frac{d\boldsymbol{\kappa'}_{\perp}}{d\varphi}\cdot\frac{d\mathbf{n}}{d\varphi}
=-\left(\sinh\xi\sqrt{\kappa_{\perp}^{'2}+\frac{3}{2}\frac{w_{\circ}}
{\tilde{\rho}^2_{\circ}}\tilde{\rho}}~\mp\cosh\xi\sqrt{\frac{w_{\circ}}
{2\tilde{\rho}^2_{\circ}}\tilde{\rho}}~\right)|d\mathbf{n}/d\varphi|^2~.
\end{equation}
Thus we should solve the system of equations (75), (\ref{77}) and
(\ref{78}) provided that $\mathbf{n(\varphi)}$ is a known function.

If non of Eqs.~(\ref{77}) and (\ref{78}) vanishes one can divide
them by each other and after some simple vector calculations obtain
\begin{equation}\label{79}
\frac{d\boldsymbol{\kappa'}_{\perp}}{d\varphi}\cdot\left[\frac{d\mathbf{n}}{d\varphi}
\times\left(\frac{d\mathbf{n}}{d\varphi}\times\boldsymbol{\kappa'}_{\perp}\right)\right]
=0~,
\end{equation}
which gives
\begin{equation}\label{80}
\frac{d\boldsymbol{\kappa'}_{\perp}}{d\varphi}=\mathbf{Z}(\varphi)\times
\left[\frac{d\mathbf{n}}{d\varphi}
\times\left(\frac{d\mathbf{n}}{d\varphi}\times\boldsymbol{\kappa'}_{\perp}\right)\right]~,
\end{equation}
where $\mathbf{Z}(\varphi)$ is an arbitrary continuous function. We
will not go further in this way but alternatively seek fore more
simple solutions. If we assume
$\frac{d\boldsymbol{\kappa'}_{\perp}}{d\varphi}=0$ then it is
possible to show after some calculations that this is not a
consistent solution for the system of equations (75), (\ref{77}) and
(\ref{78}). However, a consistent simple solution is found under the
assumption
\begin{equation}\label{81}
\frac{d\boldsymbol{\kappa'}_{\perp}}{d\varphi}\cdot\mathbf{n}=
-\boldsymbol{\kappa'}_{\perp}\cdot\frac{d\mathbf{n}}{d\varphi}=0~.
\end{equation}
This condition with the help of (\ref{77}) gives
\begin{equation}\label{82}
\kappa'_{\perp}=\texttt{const}\equiv \bar{\kappa}'_{\perp}~.
\end{equation}
By the use of (\ref{81}) and (\ref{82}) since $|\tanh\xi|<1$ we find
a differential equation relating $\tilde{\rho}$ to $\xi$ whose
solution is
\begin{equation}\label{83}
\tilde{\rho}=\frac{\bar{\kappa}^{'2}_{\perp}\tilde{\rho}_{\circ}^2}{3w_{\circ}}
\left\{\cosh\left[\pm\frac{2}{\sqrt{3}}(\xi
-\xi_{\circ})+\cosh^{-1}\left(1+\frac{3w_{\circ}}
{\bar{\kappa}^{'2}_{\perp}\tilde{\rho}_{\circ}}\right)\right]-1\right\},
\end{equation}
where the upper (positive) sign refers to the forward and the lower
(negative) sign denotes the backward sound wave. Now we should find
$\boldsymbol{\kappa'}_{\perp}$. It is clear from (\ref{81}) and
(\ref{82}) that $\frac{d\boldsymbol{\kappa'}_{\perp}}{d\varphi}$ is
perpendicular to both $\boldsymbol{\kappa'}_{\perp}$ and
$\mathbf{n}$ thus $\frac{d\boldsymbol{\kappa'}_{\perp}}{d\varphi}$
is parallel to $\mathbf{n}\times\boldsymbol{\kappa'}_{\perp}$. On
the other hand the identity
$\mathbf{n}\cdot\frac{d\mathbf{n}}{d\varphi}=0$ and Eq.~(\ref{81})
imply that $\frac{d\mathbf{n}}{d\varphi}$ is also parallel to
$\mathbf{n}\times\boldsymbol{\kappa'}_{\perp}$. Therefore we
conclude that
\[
\frac{d\boldsymbol{\kappa'}_{\perp}}{d\varphi}=
\pi(\varphi)\frac{d\mathbf{n}}{d\varphi}~,
\]
or equivalently
\begin{equation}\label{84}
\boldsymbol{\kappa'}_{\perp}(\varphi)=\int_{\varphi_{\circ}}^{\varphi}
\pi(\varphi')\frac{d\mathbf{n}(\varphi')}{d\varphi'}d\varphi'
+\boldsymbol{\kappa'}_{\perp}(\varphi_{\circ})~,
\end{equation}
where $\pi(\varphi')$ is an arbitrary nonzero continuous scalar
function (we remember that
$\frac{d\boldsymbol{\kappa'}_{\perp}}{d\varphi}$ can not vanish).
Finally we substitute (\ref{84}) and (\ref{82}) into (\ref{78}) to
obtain
\begin{equation}\label{85}
\left(\sinh\xi\sqrt{\kappa_{\perp}^{'2}+\frac{3}{2}\frac{w_{\circ}}
{\tilde{\rho}^2_{\circ}}\tilde{\rho}}~\mp\cosh\xi\sqrt{\frac{w_{\circ}}
{2\tilde{\rho}^2_{\circ}}\tilde{\rho}}~\right)=-\pi(\varphi)~.
\end{equation}
Equations (\ref{83}) and (\ref{85}) are used to express both
$\tilde{\rho}$ and $\xi$ as functions of $\varphi$ and this means
that the problem is formally solved. Substituting of all physical
quantities obtained above into the Lorentz transformation (\ref{54})
provides all the things in the laboratory frame.

\section{Symmetry analysis for the vortex mode equation}
Before starting this section, let us mention that from here on we
change all previous notations to quite new applications. So we
forget the meaning of all letters or symbols used in all preceding
sections and introduce new applications of them.

We consider Eq.~(\ref{28}) as a first order ODE and rewrite it in
the following form
\begin{eqnarray}
\frac{d{\bf k}}{d t}\cdot\Big({\bf n}-\frac{{\bf k}\cdot{\bf
n}}{{\bf k}^2+w}\,{\bf k}\Big)=0,\label{eq:1}
\end{eqnarray}
where $w$ is a constant, $t$ is treated as the wave phase, and ${\bf
k}=(k_1,k_2,k_3)$ and ${\bf n}=(n_1,n_2,n_3)$ are some vectors in
${\Bbb R}^3$ having the physical meaning of $\boldsymbol{\kappa}$
and unit normal vector to the wave front resp. We concern with the
latter equation to find its point and contact symmetry properties
and also give its fundamental invariants and a form of general
solutions.

It is notable here that in the mathematical structure of the simple
wave solution, it is necessary to take the unit length for ${\bf n}$
(see Eq.~(\ref{13})). However since Eq.~(\ref{eq:1}) is homogeneous
and linear with respect to ${\bf n}$, this condition is not
essential in obtaining any solution. This condition appears
important only for the compatibility of the simple wave structure.
Regarding this fact, we make our symmetry analysis in both cases of
arbitrary length and unit length for ${\bf n}$ and compare the
results with each other.

Throughout this section we assume that indices $i,j$ varies between
1 and 3. Also each index of a function implies the derivation of the
function with respect to it, unless specially stated otherwise.
%
%
\subsection{The point Symmetry of the Equation}

To find the symmetry group of Eq.~(\ref{eq:1}) by Lie infinitesimal
method, we follow the method presented in \cite{Ol}. We find
infinitesimal generators as well as the Lie algebra structure of the
symmetry group of that equation. In this subsection, we are
concerned with the action of the point transformation group.

The equation is a relation along with the variables of 1--jet space
$J^1({\Bbb R},{\Bbb R}^6)$ with (local) coordinate $(t, {\bf k},
{\bf n}, {\bf q}, {\bf p})=(t, k_i, n_j,  q_r, p_s)$
($\mbox{for}\;\;1\leq i,j,r,s\leq 3$), where this coordinate
involving an independent variable $t$ and 6 dependent variables
$k_i,n_j$ and their first derivatives $q_r,p_s$ with respect to $t$
resp.

Let ${\cal M}$ be the total space of independent and dependent
variables. The solution space of Eq.~(\ref{eq:1}), (if it exists) is
a subvariety $S_{\Delta}\subset J^1({\Bbb R},{\Bbb R}^6)$ of the
first order jet bundle of one--dimensional submanifolds of ${\cal
M}$, that is, graph of functions $k_i,n_j$, of elements
$(t,k_i(t),n_j(t))$ satisfying Eq.~(\ref{eq:1}) and the relations
$q_1=\displaystyle{\frac{\partial k_1}{\partial t}}$,
$q_2=\displaystyle{\frac{\partial k_2}{\partial t}}$,
$q_3=\displaystyle{\frac{\partial k_3}{\partial t}}$,
$p_1=\displaystyle{\frac{\partial n_1}{\partial t}}$, $p
_2=\displaystyle{\frac{\partial n_2}{\partial t}}$,
$p_3=\displaystyle{\frac{\partial n_3}{\partial t}}$ are all
fulfilled.

We define a point transformation on ${\cal M}$ with relations
\begin{eqnarray*}
\tilde{t}=\phi(t,k_i,n_j),\hspace{1cm}
\tilde{k}_r=\chi_r(t,k_i,n_j),\hspace{1cm}
\tilde{n}_s=\psi_s(t,k_i,n_j).
\end{eqnarray*}
where $\phi,\chi_r$ and $\psi_s$ are arbitrary smooth functions. Let
\begin{eqnarray}
v:=T\,\displaystyle{\frac{\partial }{\partial
t}}+\sum_{i=1}^3\Big( K_i\displaystyle{\frac{\partial }{\partial
k_i}}+ N_i\displaystyle{\frac{\partial }{\partial n_i}}\Big)
\label{eq:1-1}
\end{eqnarray}
be the general form of infinitesimal generators that signify the
Lie algebra ${\frak g}$ of the symmetry group $G$ of
Eq.~(\ref{eq:1}). In this relation, $T,K_i$ and $N_j$ are smooth
functions of variables $t,k_i$ and $n_j$. The first order
prolongation \cite{Ol} of $v$  is as follows
\begin{eqnarray*}
v^{(1)}&:=& v + \sum_i\,K_i^t\frac{\partial}{\partial q_i}+
\sum_j\,N_j^t\frac{\partial}{\partial p_j} ,\label{eq:1-2}
\end{eqnarray*}
where $K_i^t=D_t\,Q_1^i+T\,q_{i,t}$ and
$N_j^t=D_t\,Q_2^j+T\,p_{j,t}$, in which $D_t$ is the total
derivative and $Q_1^i= K_i - T\,q_i$ and $Q_2^j= N_j - T\,p_j$ are
characteristics of vector field $v$ \cite{Ol}. By applying $v^{(1)}$
on (\ref{eq:1}), we obtain the following relation
\begin{eqnarray}
&&\hspace{-1cm} \sum_{i}\,\,\Big\{\Big[K_i \Big(({\bf k}^2 + w)({\bf
k}\cdot{\bf n}+n_i)-2\,k_i\Big) + N_i({\bf k}^2 + w)({\bf k}^2 + w -
k_i^2)\Big]\,q_i   + \Big[K_{i\,t}-\sum_{j}\,(q_j\,K_{i\,k_j} \nonumber\\
&&\hspace{-1cm} + p_j\,K_{i\,n_j})
-q_i\Big(T_t-\sum_{j}\,(q_j\,T_{k_j}+p_j\,T_{n_j})\Big)
\Big]\Big(n_i({\bf k}^2 + w)-k_i({\bf k}\cdot{\bf n})\Big)\Big\} =
0, \label{eq:2}
\end{eqnarray}
whenever Eq.~(\ref{eq:1}) is satisfied. We may prescribe $t, k_i,
n_j, q_r, p_ s$ ($1\leq i,j,r,s\leq3$) arbitrarily while functions
$T$, $K_i$ and $N_j$ only depend on $t,k_i,n_j$. Thus
Eq.~(\ref{eq:2}) will be satisfied if and only if we have the
following equations:
\begin{eqnarray}
&&\hspace{-1.7cm} \sum_{i}\,\Big(n_i({\bf k}^2 + w)-k_i({\bf
k}\cdot{\bf n}) \Big)\,K_{i\,t}= 0,
\label{eq:3}\\[2mm]
&&\hspace{-1.7cm}  N_i({\bf k}^2 + w)({\bf k}^2 + w -
k_i^2)-\sum_{j}\,\Big(n_j({\bf k}^2 + w)-k_j({\bf k}\cdot{\bf
n})\Big)\,K_{j\,k_i} \nonumber\\[-2.5mm]
&&\hspace{-1.7cm} -\Big(n_i({\bf k}^2 + w)- k_i({\bf k}\cdot{\bf
n})\Big)\,T_t +K_i \Big(({\bf k}^2 + w)({\bf k}\cdot{\bf
n}+n_i)-2\,k_i\Big) = 0, \label{eq:4}\\[2mm]
&&\hspace{-1.7cm} T_{k_i}\Big(n_j({\bf k}^2 + w)-k_j({\bf
k}\cdot{\bf n}\Big)=0,   \label{eq:5}\\[2mm]
&&\hspace{-1.7cm} T_{k_i}\Big(n_j({\bf k}^2 + w)-k_j({\bf
k}\cdot{\bf n}\Big)+T_{k_j}\Big(n_i({\bf k}^2 + w)- k_i({\bf
k}\cdot{\bf n}\Big)=0, \label{eq:5-1}\\[2mm]
&&\hspace{-1.7cm}  \sum_{j} \Big(n_j({\bf k}^2 + w)-k_j({\bf
k}\cdot{\bf n}\Big)\,K_{j\,n_i}=0.\label{eq:6}
\end{eqnarray}
These equations are called the determining equations. From
Eqs.~(\ref{eq:4}) for each $i$ we have
\begin{eqnarray}
N_i &=&({\bf k}^2 + w - k_i^2)^{-1}\Big\{\Big(2\,k_i({\bf k}^2 +
w)-({\bf k}{\bf n}+n_i)\Big)\,K_i + \sum_j\, \Big(n_j({\bf k}^2 +
w)-k_j({\bf k}\cdot{\bf n}) \Big)\,K_{j\,k_i} \nonumber \\[-2mm]
&& + \Big(n_i({\bf k}^2 + w)-k_i({\bf k}\cdot{\bf n})
\Big)\,T_t\Big\}. \label{eq:7}
\end{eqnarray}
Since ${\bf n}\neq 0$, without loss of generality, one may assume
that $n_1\neq 0$. Also since ${\bf k}^2+w \neq 0$ so by
Eqs.~(\ref{eq:5}) and (\ref{eq:5-1}) we conclude that $T$ just
depends on $t$:
\begin{eqnarray}
T= T(t). \label{eq:8}
\end{eqnarray}
By solving Eq.~(\ref{eq:3}) with respect to $t$, we deduce the
following relation of $K_i$~s
\begin{eqnarray}
\sum_{i}\,\Big(n_i({\bf k}^2 + w)-k_i({\bf k}\cdot{\bf n})
\Big)\,K_i= 0. \label{eq:9}
\end{eqnarray}
After differentiating of the latter equation with respect to $n_j$
when we apply Eqs.~(\ref{eq:6}) we lead to the following relations
\begin{eqnarray*}
&& ({\bf k}^2 + w - k_i^2)\,K_i - \sum_{j\neq i}k_i\,k_j\,K_j=0.
\end{eqnarray*}
These relations suggest the general forms of $K_1,K_2$ and $K_3$
as follows
\begin{eqnarray}
K_1=K_2=K_3=0. \label{eq:10}
\end{eqnarray}
By applying (\ref{eq:10}) on relations (\ref{eq:7}) for different
values of $i$, the forms of $N_i$~s are also achieved:
\begin{eqnarray}
N_i &=&({\bf k}^2 + w - k_i^2)^{-1}(n_i({\bf k}^2 + w)-k_i({\bf
k}\cdot{\bf n}))\,T_t. \label{eq:11}
\end{eqnarray}
Finally, the general form of infinitesimal generators as elements of
point symmetry algebra of Eq.~(\ref{eq:1}), which we call them {\it
point infinitesimal generators}, for arbitrary functions $T$ is as
follows
\begin{eqnarray}
v = v_T := T\,\frac{\partial}{\partial t}+ \sum_{i=1}^3\Big\{({\bf
k}^2 + w - k_i^2)^{-1}(n_i({\bf k}^2 + w)-k_i({\bf k}\cdot{\bf
n}))\,T_t\Big\}\frac{\partial}{\partial n_i}.\label{eq:12}
\end{eqnarray}
The Lie bracket (commutator) of every two vector fields in the form
of (\ref{eq:12}) straightforwardly is an infinitesimal operator in
the same form of them. More explicitly, the commutator of operators
$v_T$ and $v_{\overline{T}}$ is vector field $v_{T\,\overline{T}_t -
T_t\,\overline{T}}$. Hence, the Lie algebra ${\frak g}=\langle v_T
\rangle$ of point symmetry group $G$, when $T$ is an arbitrary
smooth function which depends on $t$, is a Lie algebra.
\paragraph{Theorem 1.}{\it The set of all point infinitesimal generators in the form of
(\ref{eq:12}) is an infinite--dimensional Lie algebra of equation
(\ref{eq:1}) for arbitrary ${\bf n}$ (not necessarily unit).}\\

According to theorem 2.74 of \cite{Ol}, the invariants
$u=I(t,k_1,k_2,k_3,n_1,n_2,n_3)$ of one--parameter group with
infinitesimal generators in the form of (\ref{eq:12}) satisfy the
linear homogeneous partial differential equations of first order:
\begin{eqnarray*}
v[I]=0.
\end{eqnarray*}
The solutions of the latter, are found by the method of
characteristics (See \cite{Ol} and \cite{Ib} for details). So we can
replace the above equation by the following characteristic system of
ordinary differential equations
\begin{eqnarray}
\frac{dt}{T}=\frac{dk_1}{K_1}=\frac{dk_2}{K_2}=\frac{dk_3}{K_3}
=\frac{dn_1}{N_1}=\frac{dn_2}{N_2}=\frac{dn_3}{N_3}.\label{eq:13}
\end{eqnarray}
By solving Eqs.~(\ref{eq:13}) of the differential generator
(\ref{eq:12}), we (locally) find the following general solutions
\begin{eqnarray}
&&\hspace{-0.7cm} I_{\alpha}(t,{\bf k},{\bf n}) =
k_{\alpha}=d_{\alpha},
\hspace{2cm}\mbox{for} \hspace{0.2cm} \alpha=1,2,3 \nonumber\\[-2.5mm]
&& \label{eq:14}\\[-2.5mm]
&&\hspace{-0.7cm} I_{\beta+3}(t,{\bf k},{\bf n}) =
\frac{1}{T}\,\Big\{n_{\beta}({\bf k}^2+ w -
k_{\beta}^2)-k_{\beta}({\bf k}\cdot{\bf n})\Big\}= d_{\beta + 3},
\hspace{0.5cm}\mbox{for} \hspace{0.2cm} \beta=1,2,3. \nonumber
\nonumber
\end{eqnarray}
where $d_{\alpha}$ and $d_{\beta}$ are constants. The functions
$I_1,I_2,\cdots,I_6$ form a complete set of functionally independent
invariants of one--parameter group generated by (\ref{eq:12}) (see
\cite{Ol}).\par
Similar to the theorem of section 4.3.3 of \cite{Ib}, the derived
invariants (\ref{eq:14}) as independent first integrals of the
characteristic system of the infinitesimal generator (\ref{eq:12})
provide the general solution $$S(t,{\bf k},{\bf n}):=\mu(I_1(t,{\bf
k},{\bf n}),I_2(t,{\bf k},{\bf n}),\cdots,I_6(t,{\bf k},{\bf n})),$$
with an arbitrary function $\mu$, which satisfies in the equation
$v[\mu]=0$. This theorem can be extended for each finite set of
independent first integrals (invariants) of characteristic system
provided with an infinitesimal generator.\par
In the following, we give some examples provided with different
selections of coefficients of Eq.~(\ref{eq:12}) to show the method
explicitly. We assumed that each appeared coefficient of vector
fields is nonzero.
\paragraph{Example 1.} If we assume that $T=1$ , then the infinitesimal operator (\ref{eq:12}) reduces to
the following vector field $v_1=\frac{\partial}{\partial t}$ and the
group transformations (or flows) for the parameter $s$ are
expressible as $(t,{\bf k},{\bf n})\rightarrow (t+s,{\bf k},{\bf
n})$ which form the (local) symmetry group of $v_1$.\par
The derived invariants in this case will be as follows
\begin{eqnarray*}
I_{\alpha}=k_{\alpha},\hspace{1cm} I_{\beta+3}=n_{\beta}({\bf
k}^2+w-k_j^2)-k_{\beta}({\bf k}\cdot{\bf n}),\hspace{1cm}
\mbox{for}\:\: \alpha,\beta=1,2,3.
\end{eqnarray*}
Therefore the general solution corresponding to $v_1$ when $\mu$ is
an arbitrary function, will be $$S(t,{\bf k},{\bf n})= \mu\Big({\bf
k}\,,\,{\bf n}({\bf k}^2+ w - k_1^2)-{\bf k}({\bf k}\cdot{\bf
n})\Big).$$
\paragraph{Example 2.} Let $T=t$, then
the infinitesimal generator is
\begin{eqnarray*}
v_2=t\,\frac{\partial}{\partial t} + \sum_{j=1}^3 \,({\bf k}^2 + w
- k_i^2)^{-1}(n_i({\bf k}^2 + w)-k_i({\bf k}\cdot{\bf n}))
\frac{\partial}{\partial n_j},
\end{eqnarray*}
Then, the flows of $v_2$ for various values of parameter $s$ are
$$(t,k_i,n_j)\rightarrow\Big(t\,e^s,k_i,({\bf k}^2 + w - k_j^2)^{-1}\Big\{(n_j({\bf k}^2+w)-k_j({\bf k}\cdot{\bf n}))\,e^s
+ ({\bf k}\cdot{\bf n}-n_j\,k_j)\,k_j\Big\}\Big).$$ Also, we have
the below invariants
\begin{eqnarray*}
I_{\alpha}=k_{\alpha},\hspace{1cm}
I_{\beta+3}=\frac{1}{t}\Big(n_{\beta}({\bf
k}^2+w-k_j^2)-k_{\beta}({\bf k}\cdot{\bf n})\Big),\hspace{1cm}
\mbox{for}\:\: \alpha,\beta=1,2,3,
\end{eqnarray*}
whenever defined and the general solution of Eq.~(\ref{eq:1}) as
$$S(t,{\bf k},{\bf n}) = \mu\Big({\bf
k}\,,\,{\displaystyle\frac{1}{t}}\Big({\bf n}({\bf
k}^2+w-k_j^2)-{\bf k}({\bf k}\cdot{\bf n})\Big)\Big),$$ where $\mu$
is an arbitrary function.
\paragraph{Example 3.} In the case $T=e^t$ the infinitesimal generator (\ref{eq:12}) changes
to
\begin{eqnarray*}
v_3 = e^t\Big\{\frac{\partial}{\partial t} + \sum_{j=1}^3 ({\bf
k}^2 + w - k_i^2)^{-1}(n_i({\bf k}^2 + w)-k_i({\bf k}\cdot{\bf
n}))\,\frac{\partial}{\partial n_j}\Big\},
\end{eqnarray*}
with group transformations of the parameter $s$ transforming
$(t,k_i,n_j)$ to
\begin{eqnarray*}
&&\hspace{-0.7cm}P(s)=\left(\ln\,\{e^t(1-s\,e^t)^{-1}\}\,,\,k_i\,,\,k_j\,s\,e^t\,(1-s\,e^t)^{-1}\Big(n_j-({\bf
k}\cdot{\bf n}-n_jk_j)({\bf k}^2+w-k_j^2)^{-1}\Big) \right),
\end{eqnarray*}
wherever defined. Independent invariants are
\begin{eqnarray*}
I_{\alpha}=k_{\alpha},\hspace{1cm}
I_{\beta+3}=e^{-t}\Big(n_{\beta}({\bf k}^2+w-k_j^2)-k_{\beta}({\bf
k}\cdot{\bf n})\Big),\hspace{1cm} \mbox{for}\:\:
\alpha,\beta=1,2,3,
\end{eqnarray*}
and hence the general solution of (\ref{eq:1}) with respect to
infinitesimal operator $v_3$ is an arbitrary function of these
invariants. Indeed, if $u=f(t,k_i,n_j)$ be a solution of
Eq.~(\ref{eq:1}) then so is $u=f(P(s))$ for each $s$.

When we assume that ${\bf n}$ to be of unit length, then by the
action of $v^{(1)}$ (the first prolongation of the general form
(\ref{eq:1-1}) of infinitesimal generator $v$) on relation
$n_1^2+n_2^2+n_3^2=1$ we tend to the following equation
\begin{eqnarray*}
n_1\,N_1 +n_2\,N_2 + n_3\,N_3=0.
\end{eqnarray*}
The last equation along with the deduced form of $N_i$~s in
(\ref{eq:11}) imply that $T_t=0$. Hence $T=c$ for arbitrary constant
$c$ and for each $i$, $N_i=0$. Therefore, the form of infinitesimal
generators reduces from relation (\ref{eq:12}) to the below
expression
\begin{eqnarray*}
v=\frac{\partial}{\partial t}.
\end{eqnarray*}
\paragraph{Theorem 2.}{\it The point Lie algebra of equation
(\ref{eq:1}) when ${\bf n}$ is a unit normal vector to the wave
front is ${\frak g}=\langle \frac{\partial}{\partial t}\rangle$
isomorphic to the Lie algebra ${\Bbb R}$. Therefore the point
symmetry group of the equation with this additional condition is
the group of phase translations.}

\subsection{The Contact Symmetry of the Equation}

In continuation, we change the group action and find symmetry group
and invariants of Eq.~(\ref{eq:1}) up to the contact transformation
groups. According to B\"{a}cklund theorem \cite{Ol}, if the number
of dependent variables be greater than one (like our problem), then
each contact transformation is the prolongation of a point
transformation. In this subsection, we earn the structure of
infinitesimal generators of contact transformations. The normal
vector to wave front, ${\bf n}$, is assumed to be either arbitrary
(not necessarily unit) or of unit length and then we find the
contact symmetry properties of the Eq.~(\ref{eq:1}) in these two
situations.

We suppose that the general form of a contact transformation be as
following
\begin{eqnarray*}\begin{array}{lll}
\widetilde{t}=\phi(t,k_i,n_j,q_r,p_s),&
\widetilde{k}_l=\chi_l(t,k_i,n_j,q_r,p_s),&
\widetilde{n}_m=\psi_m(t,k_i,n_j,q_r,p_s),\\[2mm]
\widetilde{q}_n=\eta_n(t,k_i,n_j,q_r,p_s),&
\widetilde{p}_u=\zeta_u(t,k_i,n_j,q_r,p_s),&
\end{array}\end{eqnarray*}
where $i,j,l,m,n$ and $u$ varies between 1 and 6; and
$\phi,\chi_l,\psi_m,\eta_n$ and $\zeta_u$ are arbitrary smooth
functions.
In this case of group action, an infinitesimal generator which is
a vector field in $J^1({\Bbb R},{\Bbb R}^6)$, has the following
general form
\begin{eqnarray}
v:=T\,\frac{\partial }{\partial t}+\sum_{i=1}^3
\{K_i\frac{\partial }{\partial k_i}+ N_i\frac{\partial }{\partial
n_i}+ Q_i\frac{\partial }{\partial q_i}+ P_i\frac{\partial
}{\partial p_i}\}, \label{eq:16}
\end{eqnarray}
for arbitrary smooth functions $T, K_l, N_m, Q_m, P_u$ ($l=1,2$
and $1\leq m, n, u\leq 3$). \par
Since our computations are done in 1--jet space, so we do not need
to lift $v$ to higher jet spaces and hence we act $v$ (itself) on
the Eq.~(\ref{eq:1}), then we find the following relation
\begin{eqnarray*}
&&\sum_{i=1}^3\,\Big\{ N_i\,q_i({\bf k}^2+w)({\bf k}^2+w-k_i^2)+
Q_i({\bf k}^2+w)(n_i({\bf k}^2+w)- k_i({\bf
k}\cdot{\bf n})) \\
&& - K_i\,q_i(n_i({\bf k}^2+w)+({\bf k}\cdot{\bf n})({\bf
k}^2+w-2\,k_i))\Big\} =0.
\end{eqnarray*}
Since ${\bf n}\neq 0$, so without less of generality, we can
suppose that $n_1\neq 0$, then the solution to this equation for
would be
\begin{eqnarray*}
Q_1 &=& ({\bf k}^2+w)^{-1}(n_1({\bf k}^2+w)- k_1({\bf k}\cdot{\bf
n}))^{-1}\Big\{\sum_{i=1}^3\,\Big(K_i\,q_i(n_i({\bf
k}^2+w)+({\bf k}\cdot{\bf n})({\bf k}^2+w-2\,k_i))\\[-2mm]
&& -N_i\,q_i({\bf k}^2+w)({\bf k}^2+w-k_i^2) \Big)-
\sum_{j=2}^3\,Q_i({\bf k}^2+w)(n_i({\bf k}^2+w)- k_i({\bf
k}\cdot{\bf n}))\Big\}.
\end{eqnarray*}
Therefore, the infinitesimal generator which we call it as {\it
contact infinitesimal generator} is in the following form
\begin{eqnarray}
v &=& T\,\frac{\partial }{\partial t} + \sum_i\,
K_i\Big(\frac{\partial }{\partial k_i}+\frac{n_i({\bf
k}^2+w)+({\bf k}\cdot{\bf n})({\bf k}^2+w-2\,k_i))}{({\bf
k}^2+w)(n_1({\bf k}^2+w)- k_1({\bf k}\cdot{\bf
n}))}\,q_i\,\frac{\partial }{\partial q_1}\Big) \nonumber\\
&&  + \sum_i\, N_i\Big(\frac{\partial }{\partial n_i}-\frac{{\bf
k}^2+w-k_i^2))}{n_1({\bf k}^2+w)- k_1({\bf k}\cdot{\bf
n})}\,q_i\,\frac{\partial }{\partial q_1}\Big) + \sum_i\,P_i\,\frac{\partial }{\partial p_i} \label{eq:17}\\
&& + \sum_{j=2,3}\, Q_j\Big(\frac{\partial }{\partial
q_j}-\frac{n_j({\bf k}^2+w) + k_j({\bf k}\cdot{\bf n})}{n_1({\bf
k}^2+w)- k_1({\bf k}\cdot{\bf n})}\,\frac{\partial }{\partial
q_1}\Big). \nonumber
\end{eqnarray}
One may divide the latter form to the following vector fields, to
consist a basis for Lie algebra ${\frak g}=\langle v \rangle$ of
contact symmetry group $G$:
\begin{eqnarray}
&&\hspace{-1cm} v_T = T\,\displaystyle{\frac{\partial }{\partial
t}}, \hspace{2.35cm}  v_{K_i} = K_i\Big(\frac{\partial }{\partial
k_i}+\frac{n_i({\bf k}^2+w)+({\bf k}\cdot{\bf n})({\bf
k}^2+w-2\,k_i))}{({\bf k}^2+w)(n_1({\bf k}^2+w)- k_1({\bf
k}\cdot{\bf n}))}\,q_i\,\frac{\partial }{\partial q_1}\Big), \nonumber \\
&&\hspace{-1cm} v_{P_i} = P_i\,\frac{\partial }{\partial p_i},
\hspace{2cm} v_{N_i} = N_i\Big(\frac{\partial }{\partial
n_i}-\frac{{\bf k}^2+w-k_i^2))}{n_1({\bf k}^2+w)- k_1({\bf
k}\cdot{\bf
n})}\,q_i\,\frac{\partial }{\partial q_1}\Big), \label{eq:18}\\
&&\hspace{-1cm} v_{Q_j} = Q_j\Big(\frac{\partial }{\partial
q_j}-\frac{n_j({\bf k}^2+w) + k_j({\bf k}\cdot{\bf n})}{n_1({\bf
k}^2+w)- k_1({\bf k}\cdot{\bf n})}\,\frac{\partial }{\partial
q_1}\Big), \nonumber
\end{eqnarray}
where $1\leq i \leq 3$ and $2\leq j \leq 3$. The commutator of every
two of vector fields (\ref{eq:18}) is a linear combination of two
operators (\ref{eq:18}) which are generally in the form of those two
operators again. Thus these vector fields construct a basis for the
Lie algebra ${\frak g}$ of the contact symmetry group $G$. The
commutator table is given in Table \ref{table:2-1} for
$1\leq\alpha,\beta,\eta\leq 3$ and $2\leq\gamma\leq 3$. In this
table, when the commutator of two vector fields is generally in the
same form of some vector fields in (\ref{eq:18}), then we have used
those general forms again to show the results of commutators.

\begin{table}
\begin{eqnarray*}
&&\mbox{Table 5.2} \\  && \mbox{The
commutators table provided by contact  symmetry.}\\
&&\begin{array}{l|c c c c c} \hline\hline
      & v_T & v_{K_{\alpha}} & v_{N_{\beta}} & v_{Q_{\gamma}} & v_{P_{\eta}}   \\ \hline
  v_T & 0   & v_T + v_{K_{\alpha}}  & v_T + v_{N_{\beta}}& v_T + v_{Q_{\gamma}}& v_T + v_{P_{\eta}}  \\[1mm]
  v_{K_{\alpha}}&-v_T-v_{K_{\alpha}}&0 & v_{K_{\alpha}}+v_{N_{\beta}} & v_{K_{\alpha}}+v_{Q_{\beta}} & v_{K_{\alpha}} + v_{P_{\eta}}  \\[1mm]
  v_{N_{\beta}} &-v_T-v_{N_{\beta}} &-v_{K_{\alpha}}-v_{N_{\beta}} &0 &v_{N_{\beta}}+v_{Q_{\gamma}} &v_{N_{\beta}}+v_{P_{\eta}}     \\[1mm]
  v_{Q_{\gamma}}&-v_T-v_{Q_{\gamma}}&-v_{K_{\alpha}}-v_{Q_{\gamma}}&-v_{N_{\beta}}-v_{Q_{\gamma}}& 0 & v_{Q_{\gamma}}+v_{P_{\eta}}     \\[1mm]
  v_{P_{\eta}}  &-v_T-v_{P_{\eta}}  &-v_{K_{\alpha}}-v_{P_{\eta}}  &-v_{N_{\beta}}-v_{Q_{\eta}}  &-v_{Q_{\gamma}}-v_{P_{\eta}} &0 \\[1mm]
    \hline\hline
 \end{array}\end{eqnarray*}
 \label{table:2-1}
\end{table}
\paragraph{Theorem 3.} {\it The contact symmetry group of
equation (\ref{eq:1}), is an infinite--dimensional Lie algebra
generated by the contact infinitesimal operators (\ref{eq:18})
with the commutator table \ref{table:2-1}.}\\

One may repeat the above process for the problem of finding contact
Lie algebra of Eq.~(\ref{eq:1}) with the supplementary condition of
${\bf n}$ to be unit. In this case there is another condition
$v[\mbox{Eq.~(\ref{eq:1})}]=2(n_1\,N_1+n_2\,N_2+n_3\,N_3)=0$ by the
action of a contact infinitesimal generator on Eq.~(\ref{eq:1})
which must be added to other relations. Since ${\bf n}\neq 0$, so we
can suppose $n_1\neq 0$ and
$N_1=-\frac{n_2}{n_1}\,N_2-\frac{n_3}{n_1}\,N_3$. Finally we can say
\paragraph{Theorem 4.} {\it The contact symmetry group of
equation (\ref{eq:1}) consisting of unit normal wave front, is an
infinite--dimensional Lie algebra and its Lie algebra is generated
by the contact infinitesimal operators (\ref{eq:17}) when we replace
the coefficient $N_1$ by
$-\frac{n_2}{n_1}\,N_2-\frac{n_3}{n_1}\,N_3$. The commutator table
of these vector fields is in the form of Table \ref{table:2-1} when
we eliminate the row and column corresponding
to $v_{N_1}$ and then change $v_{N_2}$ and $v_{N_3}$ to the new forms.}\\

\section{Summary and conclusions}

Towards a deeper understanding of the mysterious behavior of
hydrodynamical equations it is necessary to look for various exact
solutions. Among these solutions simple waves and multi--waves are
the best for compressible flows up to present. These solutions show
explicitly as a special case that how it is possible that smooth
initial conditions convert to some discontinuities and singularities
in future times. Thus, there is a significant hope that by a
detailed and deep analysis of these waves one may find more general
statements about the appearance of any non-smoothness regarding
smooth initial conditions.

In the present work a multidimensional version of simple waves
introduced in References \cite{Boillat} and \cite{Webb2} were
employed for fully relativistic fluids and plasmas. Each wave
front is a plane traveling with its own phase velocity vector. The
intersection of different wave fronts is forbidden in the domain
of the solution. Also at each instant of time there is a surface
as the boundary between the two regions, the region of the
validity of the solution and the forbidden region where the
solution does not exist. This boundary generally moves and changes
in the course of time.

Similar to the nonrelativistic case\cite{Webb2} three essential
modes were found, namely vortex, entropy and sound modes. Each
mode suggests a wide variety of solutions while only very simple
typical solutions were presented as some illustrations of the
method of solving. Vortex and entropy modes were solved both in
the laboratory and the wave frame. But, due to the high complexity
of sound modes we studied them only in the wave frame. Further, as
a special physically valid example we considered the
thermodynamically state equation at ultra-relativistic
temperatures and obtained a complete formal solution in the wave
frame.

A symmetry analysis for the vortex mode equation (as a typical
equation) led to the finding the structure of point and contact
infinitesimal generators as well as fundamental invariants of the
equation. In addition a form of general solutions implied by these
invariants was obtained. Also we presented some examples for the
point transformation case which tend to a precise determination of
related symmetry groups. In the special case of our problem, the
contact and point symmetry group of the vortex mode equation were
both found to be infinite--dimensional Lie groups when the normal
vector to wave front is not necessarily unit. When it is of unit
length we find a one--dimensional point symmetry group while the
contact symmetry group is still infinite--dimensional. The same
procedure can be probably made for equations of other modes, namely
the entropy mode and the sound mode.



\end{document}